\documentclass[12pt]{article}

\usepackage{amsmath}
\usepackage{amssymb}
\usepackage{amsthm}
\usepackage[utf8]{inputenc}
\usepackage{graphicx}
\usepackage{algorithm}
\usepackage[noend]{algpseudocode}
\usepackage{subcaption}
\usepackage{caption} 
\usepackage{balance}
\usepackage{natbib}
\usepackage{url}
\usepackage{mathtools}
\usepackage[margin=1in]{geometry}

\usepackage{booktabs} 

\begin{document}

\title{DataPop: Knowledge Base Population using Distributed Voice Enabled Devices}
\author{
Elena Montes, Monique Shotande, Daniel Helm, Christan Grant\\
\texttt{\{elenamontes, monique.shotande, daniel.helm, cgrant\}@ou.edu}
}
\date{}



\maketitle


\begin{abstract}

   Data scientists are constantly creating methods to efficiently and accurately populate big data sets for use in large-scale applications. Many recent efforts utilize crowd-sourcing and textual interfaces. In this paper, we propose a new method of curating data; namely, creating a multi-device Amazon Alexa Skill in the form of a research trivia game. Users experience a synchronized gaming experience with other Amazon Echo users, competing against one another while filling in gaps of a connected knowledge base. This allows for full exploitation of the speed improvement offered by voice interface technology in a game-based format. 

\end{abstract}

%
%


\section{Introduction}

Knowledge Bases are increasingly becoming a structure for storing structured information.
Researchers have developed methods to gather information from the various sources to create knowledge bases~\cite{banko2007open, carlson2010toward, chen2016scalekb}.
However, these methods require significant post-processing, including probabilistic evaluation to ensure the gathered facts are correct~\cite{doan2011crowdsourcing, nakashole2014language, dong2015knowledge}.
Current methods for populating knowledge bases use textual or graphical user interfaces to enable crowd-sourced input~\cite{crescenzi2017crowdsourcing}.
This team proposes a new method that utilizes a voice interface and game format for knowledge base population.
By creating an Alexa Skill and synchronized back end, multiple Amazon Echo voice enabled devices can simultaneously interact, allowing users to play a research trivia game together. 
Given an incomplete knowledge base, natural language questions are posed to the set of users currently playing a game.
Each submitted answer is collected and the most probable answer is computed using a combination of user agreement weighted by the accuracy of each user from previous ground truth questions.
In this paper, we will motivate the need for a voice based knowledge population game (\S~\ref{sec:motivation}).
We will then, describe the architecture (\S~\ref{sec:architecture}) and game play details (\S~\ref{sec:gameplay}).
We will conclude with a description of the demo (\S~\ref{sec:demodetails}) and summary (\S~\ref{sec:summary}).

\section{Motivation}
\label{sec:motivation}

    
Traditional methods for knowledge base population rely on textual input.
Whether from mobile or desktop devices, users must type in pieces of information to be sent in for review.
This process is tedious for many users, particularly when dealing with large data sets.
Research from~\citet{ruan2016speech} finds that voice input can be up to three times faster than textual input.
New voice interface technologies used in devices, such as the Amazon Echo~\footnote{\url{https://www.amazon.com/echo}} or Google Home~\footnote{\url{https://home.google.com}}, offer an opportunity to significantly increase input speed of knowledge base population.

Many large knowledege bases are constructed by bootstrapping existing large knowledge bases~\cite{rodriguez2016sigmakb} --- Freebase~\footnote{\url{https://developers.google.com/freebase/}}, NELL~\footnote{\url{http://rtw.ml.cmu.edu/rtw/}}, OpenIE~\footnote{\url{https://nlp.stanford.edu/software/openie.html}}, ProBase~\footnote{\url{https://www.microsoft.com/en-us/research/project/probase/}}, ProbKB~\footnote{\url{https://dsr.cise.ufl.edu/projects/probkb-web-scale-probabilistic-knowledge-base/}}, and YAGO~\footnote{\url{http://www.mpi-inf.mpg.de/departments/databases-and-information-systems/research/yago-naga/yago/}} are some of the most popular bootstrapped knowledge bases.
These knowledge bases are currently popular, but many industries use the relatively smaller, topical knowledge bases such as those generated from smaller data sets or crowdsourced users imput~\cite{steiner2012seki, wang2013wisdom}.
In both large and small knowledge bases, the fusion of facts can create conflicting sets of information and the subsequently uncertain or outdated information~\cite{dong2014knowledge, grant2015challenge}.
This motivated the need and usefulness for storing the probabilities or likelihoods associated with each fact.

Many knowledge bases use paid services like Amazon Mechanical Turk for data population.
When large data sets need to be populated for large-scale applications, especially within the realm of university work or teams with a small budget, this route is simply not fiscally feasible.
With the aid of a free and open source game, for example, these teams can encourage users to assist in knowledge base population without a monetary incentive.    
The solution to the latter is to seek curating from a base of users, without a monetary incentive.
This form of basic research data entry with no discernible reward, risks user boredom or fatigue. Results may be slow or inaccurate. Gamificiation, when implemented properly, has been shown to increase user engagement and create feasible reward systems~\cite{hamari2015badges}.

\section{Architecture}
\label{sec:architecture}

This application relies on each user possessing on Amazon smart speaker device --- this may includes an Amazon Echo, Amazon Echo Dot, or an Amazon Tap. 
Each device is web enabled and contains speaker and a microphone.
These devices act as the user interface for the system.
Each user must downlaod and install a DataPop Amazon Alexa Skill.
The skill allows the user to interact with the device, voice questions, and collects responses.
A program running on our server  handles synchronization to group disparate devices/users into games and keeps track of relevant information, such as start/end timestamps and cumulative scores.
In addition, this piece keeps track of all existing users, attaching to their unique identifiers and key information such as accuracy score, the number of answered questions, and current games, if any.

To generate questions, DataPop has an interface to a knowledge base.
The program can and searches for missing or low confidence entries in the knowledge base.
The system can then generate questions from a structured format called Intents, and are sent to each Alexa system playing in the game.
The system then expects to receive answers in a particular format by specify the Slots Type.
We discuss this approach in detail in \S~\ref{sec:gameplay}.

\subsection{Hardware}

    The setup for the game interface requires multiple Amazon Echo devices, alongside a server (optional - AWS or a similar service may be used), which runs the Alexa Skill and the backend synchronized gameplay application. 

    \begin{figure}
    \includegraphics[width=\linewidth]{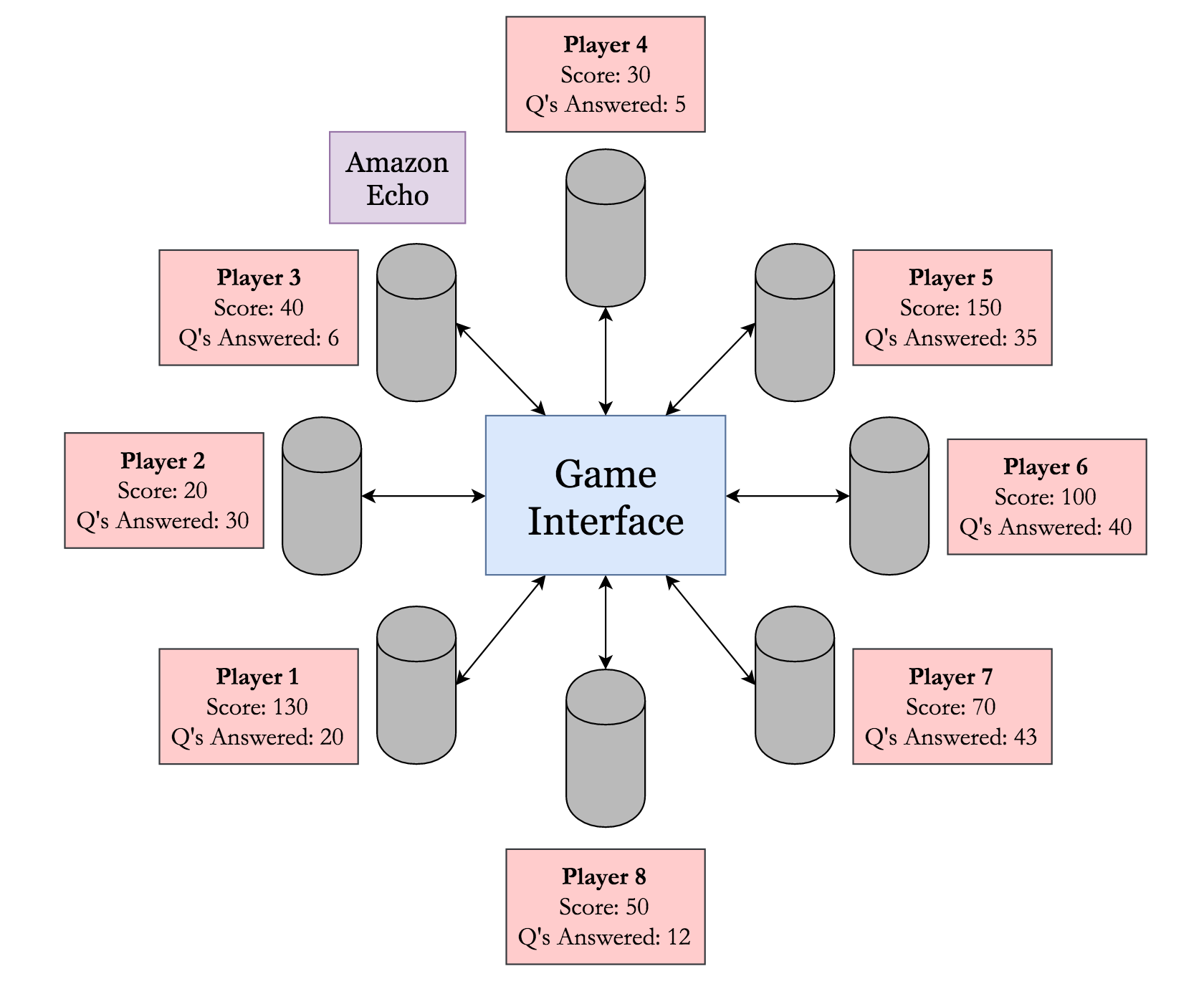}
    \caption{Hardware Setup and Attributes}
    
    \label{fig:architecture}
    \end{figure}

\subsubsection{Setup Interaction}

Self-motivated users are the most engaged and the best for populating large data sets. 
We adapt select questions for users that will most interest them
Asking relevant questions keeps users engaged over time.
Additionally, Tailoring the types of questions a user will answer to those that are relevant to their interests and preferences will increase the likelihood of accurate responses and the speed at which responses are obtained. 
    
During the Alexa Skill setup interaction (which only happens once), the user first specifies their interests upon ``signing up'' for this Skill.
The person will then answer a series of questions to gauge his/her areas of interests.
To predict user interests, we first train an five-layer neural network to accept five to ten phrases words, find the average location of those words in a word embedding.
The training process minimizes the distance between the metric spance and the labeled category.
When a set of phrases of user interest are submitted, the are fit to a set of categories, e.g., Artificial intelligence, Databases, Information Retrieval.
The categories represent a category name or predetermined cluster of a partition of the knowledge base.

\subsubsection{Further Skill Interaction}

    Once setup is completed, the skill must be invoked each time a user wishes to enter a game.
    The phrase ``Alexa, ask DataPop to let me join a game'' or similar phrases will begin this process. 
    
\subsection{Knowledge Base Preparation}


    Before the game can be played, the knowledge base chosen for gameplay must be prepared.
    The knowledge base is partitioned into categories either from clustering or using an attribute of each item in a knowledge base.
    It is instructive to picture the knowledge base as having both a row and column type.
    Columns in the knowledge base correspond to a set of related information and rows correspond to an entity.
    For a knowledge base about people, columns may refer to names, dates of births, and telephone numbers.
    
    In order to efficiently fill in the data set, some priorities must be set corresponding to to complexity of obtaining an answer in the data set.
    For example, in the person data set, dates of birth are often easier to locate than the elementary school a person attended, and thus, the former would receive a lower difficulty label.
    All the columns in the knowledge base will be ranked numerically, with zero being the most difficult and 1 being the least difficult.
        
    Question templates, or Intents, are manually created for two or more column types.
    For example, a question template, or intent may be:

    {\footnotesize
    \texttt{What is the \{Column Name\} of \{Entity Name\}?}}
   
    or
    
    {\footnotesize
    \texttt{When did \{Entity Name\} join the \{Column Name\} \{Column Value\}?}}.

\begin{figure}
    \includegraphics[width=\linewidth]{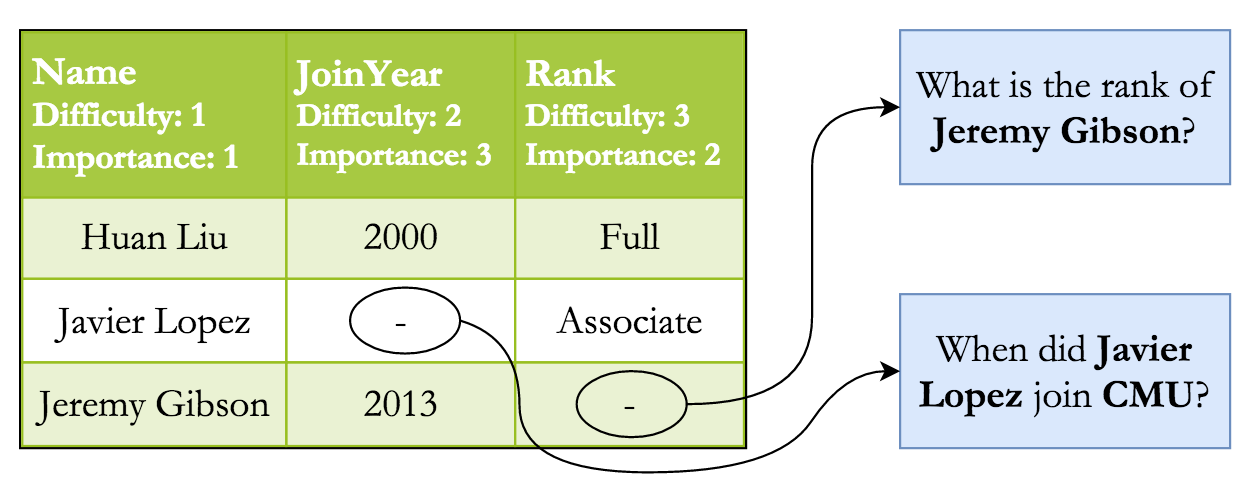}
    \caption{Example Query Generation}
    \label{fig:querygeneration}
    \end{figure}

    \texttt{Entity Name} can be a pseudo key or name in the knowledge base such as \textbf{Jeremy Gibson} or \textbf{Javier Lopez}. 
    The \texttt{Column Name} can be Slot types such as a date (in Alexa nomenclature, \textsl{Amazon.DATE} or \textsl{AMAZON.EducationalOrganization}) and the \texttt{Column Value} can be a value such as \textbf{2013} or \textbf{CMU} as seen in Figure~\ref{fig:querygeneration}.   

\section{Gameplay}
\label{sec:gameplay}

    Many aspects of the program are handled during live gameplay. 

\subsection{Game Skill Interface}
    
    Questions posed to users will be formed based on data fields that are either missing or have insufficient confidence.
    Users will have the option to either skip or answer any questions posed to them.
    Once a user answers a question, it will be recorded with a level of confidence proportional to the user's accuracy.
    Because of a limitation in the Alexa system, the user has no more than 300 seconds to respond to a question; if no response is given within this time, the query is marked as unanswered and gameplay continues.
    The user can always ask Alexa for their current score or for more time. 

\subsection{Query Creation}

    Queries are automatically generated based on a filtered-down collection of gaps detected in the knowledge base.
    From these groupings, the designers of the knowledge base can hard-code query templates.
    As the knowledge base grows, generating new templates is simple.
    Each question has an associated response type as an Amazon Slot.
    Associating a response with an type improves the accuracy of the voiced answer content. 
    
    The filtering of these gaps to choose a particular query to generate happens in stages.
    First, a collection of rows with missing information is chosen which corresponds to the topic of the current game.
    Then, if there is more than one row returned from this operation, the results are further filtered by a priority ranking.
    Each column in the data set is ranked based on its importance.
    Finally, only rows where the gap in the data occurs under the most important column, are chosen.
    If there are still multiple rows to choose from, we make a random choice and generate a query on that gap in the knowledge base. 
    
    
\subsection{Synchronization}

    Users can compete with their friends or random players in different categories, by answering the same set of questions.
    Each user has 300 seconds to answer (or skip) each question.
    Players will be informed of their opponent's answers, and overall performance, after both users have responded. 
    
\subsubsection{Player Ranking}

    Fueling a bit of competition between players will encourage continuing gameplay and long-term interest in the Skill.
    Using the method outlined in \~S~\ref{sec:accuracy}, users' accuracy scores are continually updated to help with the ranking process at the end of a game.
    Higher-ranked players will receive more rewards in the form of points, and the 'winner' of the round (with the most likely correct answer), also receives more points.

\subsubsection{Badges}

    As players accumulate points over time and work on improving their accuracy scores, they will receive audio badges that are stored on their profile. Part of the Skill's frontend allows players to check on their badge collection at any time and hear Alexa read off a listing of badges.
    This contributes to the gaming experience and works as an incentive for players to continue curating the knowledge base. 

\subsection{Accuracy Scoring}
\label{sec:accuracy}

    Two problems arise in a gamified knowledge base population: resolving conflicts and encouraging continued user participation. Giving users an accuracy score will help to resolve both issues; the former by contributing to a probabilistic determination of the most likely correct answer, and the latter by using accuracy scores to dole out badges and create some competition between game players. 
    
    Throughout gameplay, we mix in a number of predetermined queries with known --- and correct responses. A user's responses to these queries contribute to their overall accuracy score. Correct responses contribute positively, with higher difficulties gaining more points, while incorrect responses contribute zero points. 
    
    If the answer is correct, this simple formula is used to accumulate user points:
    \[\text{user}_\text{acc} = \text{user}_\text{accuracy} + \text{column}_\text{difficulty}\]
    \captionof{figure}{User Accuracy Scoring}
    

\subsection{Confidence Scoring}

    In order to account for the users providing multiple distinct results as potentially correct answers to a query, we devise a system of confidence ranking that considers the users' accuracy scores and how often an answer has been seen.
    
    The average accuracy of the users, at the time of response to the question, is used to rank each answer for any given question: 
    
    \[c_i = \frac{\sum_{j=1}^{n} ua_{ij}}{ \sum_{k=1}^{m} ua_k} \]
    
    \captionof{figure}{Confidence Scoring}
    
    Where \(c_i\) is the confidence for the ith answer; \(ua_{ij}\) is the accuracy for the jth user that provided the ith answer; \(ua_k\) is the accuracy for the user kth user that answered this question; n is the number of users that provided answer i; and m is the number of users that provided answers to the question.
    
    For example, suppose two separate answers are provided for ``What is the rank of Jeremy Gibson?'' Two users with accuracy scores of .6 and .8 provided the answer associate, and three users with accuracy scores .5, .7, and .9 provided the answer full. The confidence allotted to the associate answer will be \((.6 + .8) / (.5 + .6 + .7 + .8 + .9) = .4\); the confidence allotted to the full answer will be \((.5 + .7 + .9) / (.5 + .6 + .7 + .8 + .9) = .6\). 
    

\section{Demo Scenario}
\label{sec:demodetails}

    Two or three devices will be set up, at a distance from one another - preferably, one will be in another room while videoconferencing in. There will be one user paired to each device, and these users will go through the setup process to establish their game profiles. 
    
    A game will begin and the audience will be able to see how each user is asked the same questions and alerted to their scores at the end. If all users are not perfectly synchronized during gameplay - which is a scenario to be expected - they will receive an update notification when the game has concluded for all players. At this time, overall prizes are announced for the winner of the round.
    A laptop will be used to allow attendees to modify and explore the knowledge base.
    
    Additional features such as hearing a badge summary and help summary will also be demonstrated. 

\section{Summary}
\label{sec:summary}

    A gamified Alexa skill may be used in other areas where humans bring a unique ability to quickly interpret natural language and recognize patterns, such as in text summarization. 
    
    One weakness of this system is the limited time available for user responses. With a time constraint on answering the query Alexa gives them, the user may not be able to deliver accurate results, especially if they feel under stress.
    
    Our future plans are to expand this gamified knowledge base population method to a more generalized model that can be easily adapted to fit any existing knowledge base.

\bibliographystyle{plainnat}
\bibliography{citations} 

\balance

\end{document}